\documentclass[useAMS,usenatbib]{mn2e}

\usepackage{natbib, aas_macros}
\usepackage{ulem}
\usepackage[dvips]{graphicx}
\usepackage{wrapfig}
\graphicspath{{./figs/}}
\usepackage{color}
\usepackage{amsmath}
\usepackage{amssymb}

\newcommand{\Msun}{\rm{ M_{\odot}}}

\newcommand{\MBH}{\rm M_{\rm{BH}}}

\newcommand{\Mh}{M_{h}}

\newcommand{\Rd}{R_{\rm d}}
\newcommand{\Mvms}{M_{\rm VMS}}
\newcommand{\Mbh}{M_{\rm BH}}
\newcommand{\Trr}{t^{\rm s}_{\rm RR}}
\newcommand{\Mcl}{M_{\rm cl}}

\newcommand{\Esn}{E_{\rm SN}}
\newcommand{\Eg}{E_{\rm grav}}
\newcommand{\Rcl}{R_{\rm cl}}
\newcommand{\tcc}{t_{\rm CC}}

\title[Formation of Massive seed black holes]
{The role of stellar relaxation in the formation and evolution of the first massive black holes }
\author[Yajima et al.]
{Hidenobu Yajima$^{1, 2}$\thanks{E-mail: yajima@astr.tohoku.ac.jp (HY)} and Sadegh Khochfar$^{3}$ 
\\
$^{1}$ Frontier Research Institute for Interdisciplinary Sciences, Tohoku University, Sendai 980-8578, Japan\\
$^{2}$ Astronomical Institute, Tohoku University, Sendai 980-8578, Japan\\
$^{3}$ SUPA\thanks{Scottish Universities Physics Alliance}, 
Institute for Astronomy, University of Edinburgh, Royal Observatory, Edinburgh, EH9 3HJ, UK\\
}
\begin{document}

\date{Accepted ?; Received ??; in original form ???}

\pagerange{\pageref{firstpage}--\pageref{lastpage}} \pubyear{2016}

\maketitle

\label{firstpage}

%----------------------------------------------------------------------
%
% Abstract
%
%----------------------------------------------------------------------
\begin{abstract}
We present calculations on the formation of massive black holes with $10^5$ M$_{\odot}$ at $ z> 6$ that can be the seeds of  supermassive black holes at $z\gtrsim 6$.
Under the assumption of compact star cluster formation in merging galaxies, star clusters in haloes of $10^{8} \sim 10^{9}~\Msun$ 
can undergo rapid core-collapse leading to the formation of very massive stars (VMSs) with $\sim 1000~\Msun$ which directly collapse into black holes with similar masses. 
Star clusters in halos of $\gtrsim 10^{9}~\Msun$ experience type-II supernovae before the formation of VMSs due to long core-collapse time scales.
We also model the subsequent growth of black holes via accretion of residual stars in clusters.
2-body relaxation efficiently re-fills the loss cones of stellar orbits at larger radii and resonant relaxation at small radii is the main driver for accretion of stars onto black holes.
As a result, more than ninety percent of stars in the initial cluster are swallowed by the central black holes before $z=6$. 
Using dark matter merger trees we derive black hole mass functions at $z=6-20$. 
The mass function ranges from $10^{3}$ to $10^{5}~\Msun$ at $z \lesssim 15$. 
Major merging of galaxies of $\gtrsim 4 \times 10^{8}~\Msun$ at $z \sim 20$ successfully leads to the formation of $ \gtrsim 10^{5}~\Msun$ BHs by $z \gtrsim 10$ which can be the potential seeds of supermassive black holes seen today. 

\end{abstract}

%----------------------------------------------------------------------
%
% Keywords
%
%----------------------------------------------------------------------

%----------------------------------------------------------------------
%
% Section 1: Introduction
%
%----------------------------------------------------------------------
\section{Introduction}
Recent observations of high-redshift QSOs at $z > 6$ suggest supermassive black holes (SMBH) of $\Mbh \gtrsim 10^{9} ~\Msun$  form on short time scale $\lesssim \rm 1~Gyr$ \citep[e.g.,][]{Fan01a, Fan06b, Mortlock11, Kormendy13,  Wu15}.
The most distant QSOs are observed at $z = 7.1$ which corresponding to an age of the Universe of $\sim 800 ~\rm Myr$.
The formation mechanism of SMBHs on such short time scales has not been understood yet.  
Cosmological simulations show that BHs can grow up to $\sim 10^{9}~\Msun$ via Bondi-Hoyle accretion
\citep{DiMatteo08, DiMatteo12, Li07}. 
While there are uncertainties associated with the modelling of the accretion efficiency at the sub-grid level \citep{Booth09},  it appears that the  main crucial assumption is that massive black holes (MBHs) of $10^{5}~\Msun$ are seeded in haloes of $\sim 10^{10}~\Msun$ within the simulations.
These simulations indicate that once MBHs form in the first galaxies, they can grow to SMBH via gas accretion at Eddington rate by $z\sim 6$. 
The main obstacle is thus referred to forming MBHs.

One  natural scenario for the formation of massive BHs is the Eddington-limited growth of a few 100 M$_{\odot}$ stellar mass black hole via gas accretion that is the remnant of a Population III star.
However, due to stellar feedback, the gas accretion rate is significantly suppressed and far below the Eddington-limit \citep{Johnson07, Milosavljevic09a, Milosavljevic09b, Alvarez09, Park11, Park12}.
In addition, cosmological simulations show that  Population III star remnants mostly resided in low-density environments, 
hence the gas accretion rate is much smaller than the Eddington rate even without stellar feedback \citep{Alvarez09}.

Another proposed mechanism is the direct collapse of super massive stars (SMS) \citep{Rees78, Bromm03}.  If the formation of hydrogen molecules in the first galaxies is suppressed by external UV radiation in the  Lyman-Werner bands, gas clouds cannot fragment. 
As a result, the clouds cannot form low mass stars but form SMSs of $\sim 10^{5}~\Msun$ due to  high-gas accretion rates of $\sim 0.1-1.0~\Msun \; yr^{-1}$
\citep{Omukai05, Lodato06, Lodato07, Begelman06, Spaans06, Inayoshi12, Agarwal12, Agarwal13, Hosokawa13}.
This scenario is also supported by detailed numerical simulations \citep{Regan09, Regan14, Mayer10, Choi13, Latif13, Johnson14, Inayoshi14a, Agarwal14}. 
One of the main requirements of this scenario is a strong UV background radiation with $J_{21} \gtrsim 100$ in units of $10^{-21}~\rm erg \; s^{-1} \; cm^{-2}~\; Hz^{-1}$
and very low metallicity ($Z \lesssim 10^{-5}~Z_{\odot}$)  \citep{Omukai05, Omukai08, Dijkstra08c, Agarwal15, Sugimura14, Inayoshi14b}. 
Note that, however, the critical value of $J_{21}$ is still under the debate. 
It sensitively depends on the environments via, e.g., the shape of SED \citep{Sugimura14, Agarwal15, Agarwal15b}. 

Recently \citet{Mayer10} showed in high-resolution simulations that another suggested path to the formation of MBHs via major mergers of galaxies could work. 
Their simulations showed that a large amount of gas in disc galaxies falls down to the galactic centres due to 
angular momentum loss by tidal force, and massive high-density gas clumps form. 
They suggest that the gas clouds directly collapse into MBHs. 

In addition, \citet{Inayoshi15} suggested that SMSs could form in merging primordial haloes 
due to collisional dissociation of $\rm H_{2}$ in the compressed high-density regions.
The outcome of the collapse depends sensitively  on the equation of state of the gas. 
If radiative cooling quickly occurs, a collapsing gas cloud fragments and forms a dense star clusters instead \citep[e.g.,][]{Regan09, Ferrara13}.

Due to the gravothermal collapse, such dense star clusters can cause the merging of massive stars and runaway growth,
resulting in the formation of very massive stars (VMSs) of $\sim 1000~\Msun$ \citep[e.g.,][]{Freitag06}. 
 Direct numerical simulations by \citet{Portegies-Zwart02} support this view and show that compact dense star clusters cause core-collapse and make VMSs  within the typical lifetime of massive stars \citep[see also,][]{Fujii14}.
 VMSs can result in BHs with almost the same mass \citep{Heger03}. Subsequently the BHs can grow via accretion of stars with tidal disruption events \citep[e.g.,][]{Rees78}.
 \citet{Devecchi09} analytically modeled the formation of VMSs in first galaxies via disc instabilities, 
 and showed a fraction $\sim 0.05$ of first galaxies at $z \sim 10-20$ form VMSs \citep[see also][]{Devecchi12}.

 Very recently \citet{Katz15} followed the merging processes of metal enriched haloes in detailed cosmological hydrodynamics simulations
 and showed the formation of high-density gas clumps which are potential sites of dense star clusters leading to the formation of VMSs.
 In this work, we investigate the possibility of the formation of MBHs via the formation of VMSs and subsequent growth by stellar relaxation processes in merging galaxies at $z > 6$.
 
%In this work, we investigate the possibility of the formation of MBHs via  merging and tidal disruption of stars in dense star clusters, that formed in mergers of galaxies at  $z > 6$.

Our paper is organized as follows.
We describe our model in Section~\ref{sec:model}. 
In Section~\ref{sec:result}, we show the masses of VMSs as a function of halo mass, and the final BH masses, and the mass function of MBHs. 
In Section~\ref{sec:discussion}, we investigate the effects of Population III stars and the nature of metal poor globular clusters. 
Finally, in Section~\ref{sec:summary}, we summarize our main conclusions.

%in prep. \\[5.0cm]

%----------------------------------------------------------------------
%
% Section 2:  Model and Method
%
%----------------------------------------------------------------------
\section{Model}
\label{sec:model}
The starting point of our model is the formation of compact star clusters in major mergers $(M_1/M_2  < 3.5$, $M_1\geq M_2$) of gas-rich disc galaxies. It has been shown in a series of numerical simulations that during such events gas initially in a rotationally supported disc looses angular momentum and collapses toward the potential minimum of the merger remnant  
\citep{Barnes96, Springel05a, DiMatteo05, Naab06, Hopkins06, Cox06, Li07, Mayer10}. 
This gas will reach high densities on short time scales and be available to fuel star formation in the centre. 
The low amount of angular momentum facilitates the formation of compact star clusters \citep{Regan09}.
In practice, the enhancement of mass inflow and the associated starburst due to the merger depend on physical conditions such as  
e.g. the mass ratio, impact parameter and inclination \citep{Hopkins09}. 
In this work, we consider major mergers with a parameter range leading to most efficient mass inflow and associated starburst, 
i.e., $M_1/M_2  \lesssim 2$ and an impact parameter $\lesssim R_{\rm vir}$ \citep{Khochfar06}.

The progenitor disc is assumed to relate to the hosting dark matter halo via $\Rd \sim \lambda R_{\rm vir}$ \citep[e.g.,][]{Mo98},  
%
%\begin{equation}
%\Md \sim \lambda \Mh
%\end{equation}
%
where $\lambda$ and $R_{\rm vir}$ are spin parameter and  virial radius, respectively. 
%\adg{Note that, however, this argument is simplified, the disc size can be changed due to cooling, stellar feedback and tidal effect (e.g., Hopkins et al. 2009; Sales et al. 2012).}
In this work, we assume $\lambda=0.05$, however, our results are not depending on the specific choice of  $\lambda$ as we will show below. 

We model the progenitor discs of merging galaxies as a Mestel, isothermal profile, i.e., $\Sigma(r) = \Sigma_{0} (\Rd /r)$,
where $\Sigma_{0}$ and $R_{\rm d}$ are scale parameters and estimated as a function of halo properties \citep{Mo98, Devecchi10}, 
\begin{equation}
\begin{split}
\Sigma_{0} &= 70 \left( \frac{V_{\rm h}}{15~\rm km\; s^{-1}} \right) \left( \frac{f_{\rm disc}}{0.05}\right) \left( \frac{\lambda}{0.05}\right)^{-1}  ~{\rm \Msun \; pc^{-2}}, \\
\Rd &= 100 \left(  \frac{\lambda}{0.05} \right)  \left(  \frac{R_{\rm vir}}{700~\rm pc} \right)~{\rm pc}.
%\Rd = 2\sqrt{2}\left(  \frac{j_{\rm d}}{\Md} \right) \lambda R_{\rm vir}
\end{split}
\end{equation}
Here $V_{h}$ is the circular velocity, 
and $f_{\rm disc}$ is the fraction of the disc mass with respect to the total mass including bulge and dark matter. 
In this work, we assume a fiducial value of $\lambda = 0.05$ and $f_{\rm disc} = \lambda$ \citep[e.g.,][]{van-den-Bosch01}. 

The amount of inflowing gas during the merger is estimated using the prescription in \citet{Hopkins09}, which assumes that during the merger a gaseous and stellar bar develop which are out of sync and exert a torque on each other.
The resulting mass inflow of gas is then estimated by, 
\begin{equation}
%M_{\rm inf} = 2 \pi \Sigma_{0} \frac{1}{\nu_{\phi}} \frac{G M_{\rm bar}}{\Rd} \Delta \tau
M_{\rm inf} = 2 \pi \Sigma_{0} R_{\rm d} V_{h} (1 - f_{\rm gas}) f_{\rm disc} \Psi_{\rm bar} \Delta \tau,
\label{eq:minf}
\end{equation}
where %$\nu_{\phi}$ is frequency of disc rotation, and 
$f_{\rm gas} $ is the gas fraction in the disc, 
%$f_{\rm disc}$ is the fraction of the disc mass to the total mass including bulge and dark matter \citep[$\sim \lambda$:][]{van-den-Bosch01}, 
$\Psi_{\rm bar}$ is the mass fraction of stars in the bar, and
$\Delta \tau$ is the time since the merger.
We assume $f_{\rm gas}=0.9$, i.e., $10 \%$  of the disc is in stars before the merger takes place.
The mass of the resulting star cluster increases with decreasing $f_{\rm gas}$, leading to formation of more massive BHs  as we will show below. 
As suggested  in \citet{Hopkins09}, we set $\Psi_{\rm bar} = 1$  based on our choice of mass ratio, inclination and impact parameter of mergers.

%In practice, $\Psi_{\rm bar}$ can be smaller than unity depending on the major condition,s e.g., mass ratio, inclination and impact parameter}

We consider the  initial $3~\rm Myr$ after the merger, which corresponds to 
the life time of massive stars and the onset of supernovae feedback which will halt star formation in the star cluster.

%============================
%Once mass inflow occurs the inner structure of the disc will change and become much steeper in the inner parts \citep{Mineshige97}:   
% 
%\begin{equation}
%\Sigma(r) = \begin{cases}
%\Sigma_{\rm in} \left(  \frac{r}{\Rd} \right)^{-\frac{5}{3}} ~~&{\rm at}~ r \le R_{\rm tr}\\
%\Sigma_{0} \left(  \frac{r}{\Rd}\right)^{-1} ~~&{\rm at}~ r > R_{\rm tr}
%\end{cases}
%\end{equation}
%
%where $R_{\rm tr}$ is the transition radius. 
%$R_{\rm tr}$ is estimated via,
%
%\begin{equation}
%R_{\rm tr} =  \frac{M_{\rm inf}}{4 \pi \Sigma_{0}\Rd}.
%\end{equation}
%
%$\Sigma_{\rm in}$ is given by $\Sigma_{\rm in} = \Sigma_{0} \left(  \frac{R_{\rm d}}{R_{\rm tr}} \right)^{-2/3}$.
%We assume that 
%stars form in the region where the inflowing gas is stacked, 
%hence the radius of the star clusters ($R_{\rm cl}$) is $\sim R_{\rm tr}$ \citep[see also][]{Devecchi09, Devecchi10}. 
%============================

Once mass inflow occurs the inner structure of the disc will change.
However, we here focus on regions at galactic centres that are much smaller than these scales and we assume that the gas distribution is spherical.  
%\adg{Here we focus on much smaller regions at galactic centres than galactic disk scales, hence assume spherical gas distribution. }
 Prior to star formation we approximate the density profile of the central clouds with a singular isothermal profile,
\begin{equation}
\rho_{\rm gas} = \frac{c_{\rm s}^{2}}{2 \pi {\rm G} r^{2}}
\end{equation}
where $c_{\rm s}$ is the  sound speed which we approximate using  the virial temperature of the hosting halo. 
By integrating the density profile, we determine the radius of star clusters, i.e.,
$M_{\rm inf} = \int_{0}^{r_{\rm cl}} 4 \pi r^{2} \rho_{\rm gas}dr$,  hence, 
\begin{equation}
r_{\rm cl} = 2.15~{\rm pc}~ \left( \frac{c_{\rm s}}{\rm 10~km\; s^{-1}}\right)^{-2} \left( \frac{M_{\rm inf}}{10^{5}~\Msun} \right).
\end{equation}
In general, the star formation rate in such gas cloud can be parameterised by $\dot{M}_{\rm star} \sim \eta \frac{M_{\rm gas}}{t_{\rm dyn}}$
where $\eta \lesssim 0.05$ is the star formation efficiency \citep[e.g.,][]{Krumholz12}.
The dynamical time of the gas clouds in our model is typically $\lesssim 10^{4}~\rm yr$ which is much shorter than  $3~\rm Myr$.
Therefore, we assume the fiducial case of  $M_{\rm cl} \sim M_{\rm inf}$.
On the other hand, before supernovae occurs, radiative feedback may reduce the conversion efficiency from gas to stars due to photo-evaporation of gas. 
Hence, we also study the case of $M_{\rm cl} = 0.3 M_{\rm inf}$.

%============================

The star clusters undergo core-collapse, and massive stars migrate toward the centre of the star cluster. 
Recent N-body simulations show that the time scale of core-collapse is similar to that of dynamical friction \citep{Fujii14},
\begin{equation}
t_{\rm CC} \sim t_{\rm df} = \frac{1.91}{{\rm ln} (\Lambda)} \frac{r_{\rm cl}^{2} \sigma}{G m_{\rm max}}
\end{equation}
where ${\rm ln} (\Lambda)$ is the Coulomb logarithm, $\sigma$ is the three-dimensional velocity dispersion of stars and $m_{\rm max}$ is the maximum mass of stars based on the  initial stellar mass spectrum.  We assume that the stellar cluster is that of a King profile and use a Salpeter initial mass function with the mass range of $m=0.1-100~\Msun$.
Due to core-collapse, the stellar density at the centre quickly rises. 
As a result, at the centre, stars frequently collide with each other and coalesce, resulting in the formation of very massive stars (VMSs) of $\sim 1000~\Msun$ \citep{Portegies-Zwart02}.
Numerical simulations estimate the final mass of such VMS as \citep{Portegies-Zwart02}:
\begin{equation}\label{vms}
\Mvms  = m_{\rm max} + 4 \times 10^{-3} M_{\rm cl} f_{\rm c} {\rm ln}(\Lambda) {\rm ln}\left( \frac{\rm 3~Myr} {t_{\rm CC}}\right)
\end{equation}
where $M_{\rm cl}$ is the mass of the star cluster, 
 and $f_{\rm c}$ is the factor used to calibrate the analytical estimate against direct $N$-body simulations \citep[$f_{\rm c}=0.2$;][]{Portegies-Zwart02}. 

If the core-collapse time scale $t_{\rm CC}$  is longer than 3 Myr, stars will explode as type-II supernovae and we set  $\Mvms = m_{\rm max}$.
Following \citet{Portegies-Zwart02} and \citet{Fujii14}, we  use a King profile with $W_{0}=3$ for the  stellar density distribution of our model star clusters. 
If the mass of the VMSs are greater than $\sim 260~\Msun$, VMSs directly collapse to BHs \citep{Heger02}. 
%\adm{In this work, we keep the same radii for star clusters and constant mass spectrum of stars after the core-collapse. } 
After core-collapse, star clusters can shrink by a factor of a few  \citep{Fujii14}, 
massive stars tend to distribute near the centres of star clusters due to mass segregation. 
We here will use the initial state of the star cluster in our calculations, noting that the calculated growth rates will be lower limits based on the fact that we do not take into account core collapse and mass segregation in the cluster.
Subsequent disruptions of stars by the BH deplete the loss cone and drive the growth of the BH to a halt. However, the loss cone can be refilled by stars that loose angular momentum and migrate towards the centre or have a high ellipticities. 
In this work, we consider  angular momentum transport due to relaxation processes between stars. 
We will focus on two main types of relaxation processes, {\it resonant} and {\it non-resonant}.   

Two-body relaxation (non-resonant:NR) takes place over a time scales of 
%(e.g., Hopman et al. 2006), 
\citep[e.g.,][]{Binney08, Kocsis11}, 
\begin{equation}
%t_{\rm NR}(r) = A_{\rm \Lambda} \left(  \frac{\Mbh}{\Mstar} \right)^{2} \frac{P(r)}{N(<r)},
t_{\rm NR}(r)=0.34 \frac{\sigma^{3}}{G^{2} \rho m_{2} {\rm ln}(\Lambda)},
\label{eq:NR}
\end{equation}
where $\rho$ is the stellar density, and $m_{2}=<m^2>/<m>$ is the effective mass.
For a Salpeter IMF int the mass range $0.1-100~ \Msun$, $m_{2}=4.8~\Msun$.
%where \adb{$A_{\rm \Lambda} = ??$ }is a dimensionless constant including the Coulomb logarithm, $\Mstar$ is the mean stellar mass of stars for a given IMF, $P(r)=2 \pi [R^{3}/(GM_{\rm BH})]^{1/2}$, and $N$ is number of stars within radius $r$. 

The second relaxation process we consider  is resonant relaxation (RR).
Stars close to MBH move on Kepler orbits. 
These stars cause wire-like fluctuation in the gravitational potential governed by the MBH, and induce perturbations to stellar orbits. 
In particular, stars on the same plane exchange angular momentum via  scaler RR resulting in a change of ellipticities of the orbits.
As a result, stars with high ellipticity enter within the  tidal radius of the central MBH and will be disrupted by it. 
The effect of scaler RR can be suppressed by Newtonian and general relativistic precession \citep{Hopman06}, in contrast to the classical Keplarian case.

The scalar RR  time scale is estimated by \citep{Hopman06}, 
\begin{equation}
t^{\rm s}_{\rm RR}(r) = \frac{A_{\rm RR}}{N (<r)} \left(  \frac{\Mbh}{\overline{m}} \right)^{2} P^{2}(r) \left| \frac{1}{t_{\rm M}} - \frac{1}{t_{\rm GR}} \right|.
\label{eq:RR}
\end{equation}
where $A_{\rm RR}$ is a numerical factor of order unity \citep[$ \sim 3.56$;][]{Rauch96}, 
$\overline{m}$ is a mean stellar mass,
$P(r)=2 \pi [r^{3}/(GM_{\rm BH})]^{1/2}$, 
${t_{\rm M}}$ and $t_{\rm GR}$ are the Newtonian and general relativistic precession time scales. 
Following \citet{Hopman06}, we estimate the Newtonian and general relativistic precessions as follows,
$t_{\rm M}= A_{\rm M} \frac{M_{\rm BH}}{N(r) \overline{m}}P(r)$, 
and 
$t_{\rm GR} = \frac{8}{3} \left(  \frac{cJ}{4GM_{\rm BH}}\right)^{2} P(r)$, 
where $A_{\rm M} =1$ is assumed.
The shorter of the two relaxation time scales is used to estimate the loss cone refilling and  growth of BHs, 
\begin{equation}
t_{\rm relax}(r) = {\rm min}(t_{\rm NR}, \Trr).
\end{equation}

At radii where the interior mass is dominated  by the stars and not the BH we only consider $t_{\rm NR}$, because the stellar orbits are not Keplerian anymore. 
Over the relaxation time scale $T_{\rm relax}$, the angular momentum of stars is transported, 
and hence, the loss-cone is refilled, causing the growth of BHs. 
We estimate the BH mass by integrating the stellar density profile and time evolution as follows \citep[see also][]{Chen13},
\begin{equation}
\Mbh(t) = \Mbh^{\rm init} + \int_{0}^{t} dt \int_{0}^{r_{\rm cl}} \frac{4 \pi r^{2} \rho_{\rm star} (r, t)}{t_{\rm relax}(r, t)} dr.
\label{eq:mbh}
\end{equation}

Since the accretion time-scale is shorter than the relaxation time-scale of the star cluster \citep{Shiokawa15}, 
we assume that stars entering the tidal radius are instantaneously swallowed by BHs. 
Note that, however, this growth rate is an upper limit. 
In practice, some fraction of gas in stars can be unbounded when they are tidal disrupted \citep{Guillochon13}.  
%\sout{This gas will mostly settle in an accretion disc where it is subject to feedback effects from the accreting BH (Strubbe \& Quataert 2009).}% \citep{Strubbe09}}.
Even initially bound gas, that settles in an accretion disc, can be subject to feedback from the accreting BH \citep{Strubbe09}.
In addition, some stars near the central high-density region can form binaries. 
If the binaries get close to the BHs, some stars can be kicked out due to three-body interaction. 
Thus, the growth rate may be reduced by factor 2-4, 
e.g. in cases where the mass fraction of the unbounded gas to the initial stellar mass is $\sim 0.5$ and $\dot{m}_{\rm outflow} \lesssim \dot{m}_{\rm accretion}$ \citep{Strubbe09}.  

%%%%%%%%%%%%%%%%%%%%%%%%%%%%%%%%%%%%%%%

%----------------------------------------------------------------------
%
% Section 3:  Results
%
%----------------------------------------------------------------------

\section{Results}
\label{sec:result}

\subsection{Formation of very massive stars}
\begin{figure}
\begin{center}
\includegraphics[scale=0.43]{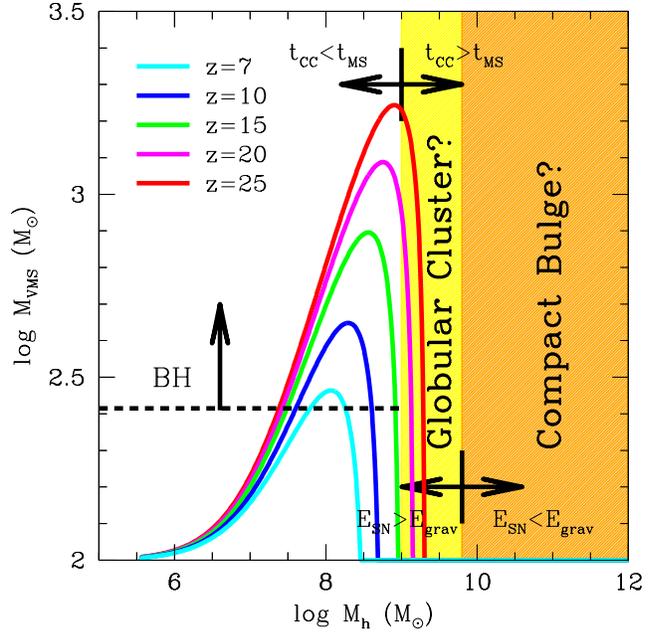}
\caption{
Mass of very massive stars in star clusters after core-collapse. 
Different colour lines represent different redshifts at which galaxies merge. 
Dashed line shows the threshold mass of stars that result in black holes at end of their life time \citep{Heger03}.
}
\label{fig:mvs}
\end{center}
\end{figure}

Figure~\ref{fig:mvs} shows the predicted mass of VMSs based on Eq. \ref{vms} as a function of the halo mass of the remnant. 
Most notably the mass of VMSs drops beyond halo masses of $M_{\rm{h}} \sim  10^{8.5-9}$. This is a direct consequence of the increasing core collapse time scale $t_{\rm{CC}}$  in larger haloes.
The difference in predicted masses of VMSs as a function of redshift is again related to the longer core-collapse time scales at low redshifts, which result in systematic lower masses. 
Once the core collapse time scale is larger than the life time of massive stars, i.e. $3$ Myr, we assume that collisions between stars become negligible during the life-time of the star, and no VMS is formed. The long relaxation times would even in the case that a remnant black hole forms, be too long to grow it efficiently. 
The mass of VMSs increases with halo mass, because the mass of star clusters increases with halo mass in our model. 
However, with  increasing cluster mass and radius, the core-collapse time becomes longer too. 
The mass growth of VMSs is logarithmic and only weak until $t_{\rm CC} \sim \rm 3~Myr$.
Haloes with $\Mh \sim 10^{9}~\Msun$ at $z \gtrsim 20$ can host VMSs of $\sim 1000~\Msun$. 
Note that, in this work, we assume low stellar metallicities ($Z \lesssim 10^{-3}~Z_{\odot}$).
If stellar metallicities are higher than $Z \gtrsim 10^{-3}~Z_{\odot}$, 
stars experience strong stellar winds and loose a large fraction of their mass before their death. 
%The maximum mass of VMSs increases with decreasing redshift from $400~\Msun$ for a halo of $10^{8.5}~\Msun$ at $z=20$ to $1400~\Msun$ for a halo of $10^{9}~\Msun$ at $z = 7$. 
The maximum mass of VMSs increases with redshift from $290~\Msun$ for a halo of $1.1\times10^{8}~\Msun$ at $z = 7$ to $1200~\Msun$ for a halo of $6.7\times 10^{8}~\Msun$ at $z=20$. 
In our model, $\Mcl$ for a specific halo mass does not depend on redshift, because in Equation~\ref{eq:minf}, 
$\Sigma_{0} \propto (1+z)^{1/2}$, $V_{h} \propto (1+z)^{1/2}$, and $R_{\rm vir} \propto (1+z)^{-1}$.
On the other hand, 
$\Rcl \propto (1+z)^{-1}$, hence star clusters become bigger with decreasing redshift. 

As a result, the core-collapse time becomes longer, and the mass of VMSs decreases. 
If the stellar metallicity is low, even $\sim 100~\Msun$ stars can directly collapse into BHs \citep{Heger03}. 
Feedback by  supernovae can suppress further gas accretion or evacuate gas from galaxies. 
%We estimate the total energy of supernovae and gravitational binding of gas as follows,
%
%\begin{equation}
%\begin{split}
%E_{\rm SN} % &\sim \int_{8~\Msun}^{40~\Msun} m \frac{dn}{dm} dm \times 10^{51}~{\rm erg}  \\
%&\sim 0.007 \left(  \frac{\Mcl}{\Msun}\right) \times 10^{51}~{\rm erg} \\
%E_{\rm grav} &\sim \frac{G M^{2}_{\rm d}}{\Rd} \\
%\frac{E_{\rm SN}}{E_{\rm grav}} &\propto \frac{R_{\rm vir}}{\Mh} \propto \Mh^{-\frac{2}{3}} (1+z)^{-1}
%\end{split}
%\end{equation}
%
%where $0.007$ is the mass fraction of massive stars causing type-II SNe given a Salpeter IMF within the mass range $0.1 - 100~\Msun$. 
%The properties of the globular clusters will be discussed in Section~\ref{sec:discussion}.

Once gas is evacuated from the disc and becomes part of the hot diffuse halo, it will take a cooling time, for it to join the disc again.
Here we roughly estimate the feedback necessary to unbind the gas from the disc. 
The feedback energy by SNe is 
$E_{\rm SN} \sim 0.007 \left(  \frac{\Mcl}{\Msun}\right) \times 10^{51}~{\rm erg}$,
where $0.007$ is %the mass fraction of massive stars causing type-II SNe given a Salpeter IMF in the mass range $0.1 - 100~\Msun$. 
the conversion factor from the star cluster's mass to number of massive stars causing type-II SNe given a Salpeter IMF in the mass range $0.1 - 100~\Msun$. 
From Equation~(\ref{eq:minf}), $E_{\rm SN} \propto \Mh$.
The disc binding energy is $E_{\rm grav} \sim \frac{G M^{2}_{\rm d}}{\Rd}  \propto \frac{\Mh^{2}}{R_{\rm vir}}$.
Then, $E_{\rm SN} / E_{\rm grav} \propto R_{\rm vir} \propto (1+z)^{-1}$. Hence gas in discs can easily be evacuated at lower redshift, due to its lower density.
The threshold mass for haloes with $E_{\rm grav} > E_{\rm SN}$ is 
$\Mh \gtrsim 1.6 \times 10^{9} \left( \frac{1+z}{11}\right)^{1.5}$.

As shown in the figure 1, haloes of $\Mh \sim 10^{7.5} - 10^{9}~\Msun$, which are in the typical mass range of the first galaxies \citep{Wise12a, Johnson13, Paardekooper13},   form compact star clusters and cause the  formation of VMSs. The VMSs of haloes with $\Mh \lesssim 10^{7.5}~\Msun$ result in pair-instability SNe \citep{Heger03}. 

%%%%%%%%figs

\subsection{Growth of BHs}
\begin{figure}
\begin{center}
\includegraphics[scale=0.43]{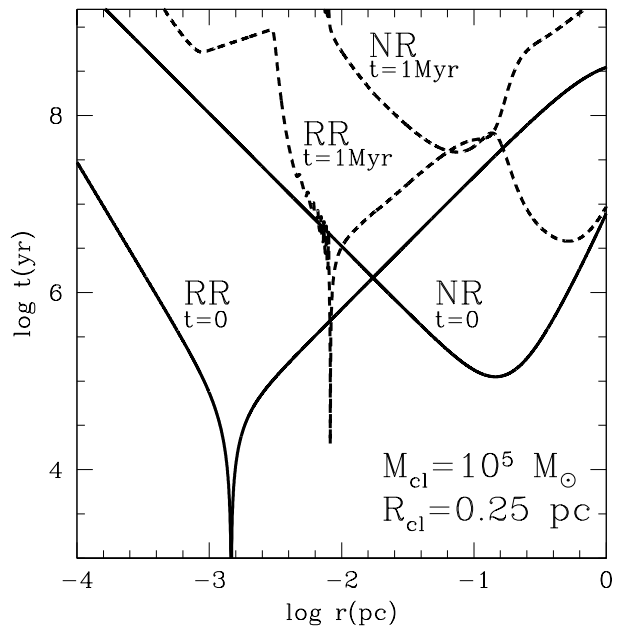}
\caption{
Time scales of stellar relaxation processes as a function of radial distance from centre of a star cluster.
NR and RR represent non-resonant relaxation (2-body) and resonant relaxation. 
Solid and dash lines show the relaxation time scale at $t=0$ and 1 Myr, respectively.
The time scales are calculated for a star cluster with the mass of $10^{5}~\Msun$ and the radius of $0.25~\rm pc$.
}
\label{fig:rrtime}
\end{center}
\end{figure}

Once a VMS of $>260~\Msun$ is formed collapse to a BH will ensue in a star cluster.  In such a system RR/NR efficiently changes the angular momentum of stars, and refills the loss cone around the BH allowing it to feed.
Figure~\ref{fig:rrtime} shows the RR and NR time scale as a function of radial distance from the centre of a star clusters. 
Due to general relativistic precession a shortest time scale exists at a specific position. 
For a star cluster with $\Mcl = 10^{5}~\Msun$ and $\Rcl = 0.25~\rm pc$ which is representative of the result of a merger between proto-galaxies with $\Mh = 3.8\times10^{8}~\Msun$ at $z=15$, 
the RR time scale becomes shorter than NR at $r \lesssim 0.02~\rm pc$. 

Hence, RR efficiently works for stellar accretion near the BH, and NR leads to the angular momentum change of stars in the outer parts.
%Then, due to the decrease of stellar density, the relaxation time scale changes. 

As the stellar density decreases with time, the relaxation time scale becomes longer. 
The RR and NR time scales at $t=1$ Myr are shown as the dash lines in the figure. 
The time scales get longer by a factor $\gtrsim 2$ from the initial state. 
The radius at which the RR time scale becomes shorter than the NR propagates outwards with time. 
In the fiducial case shown in figure~\ref{fig:rrtime}  the radius is $0.1$ pc at $t=1$ Myr.
Finally, 
depending on the RR and NR processes, the time scale for refilling of the loss cone can be shorter than $\sim 10^{8}~\rm yr$ in the clusters. 

\begin{figure}
\begin{center}
\includegraphics[scale=0.43]{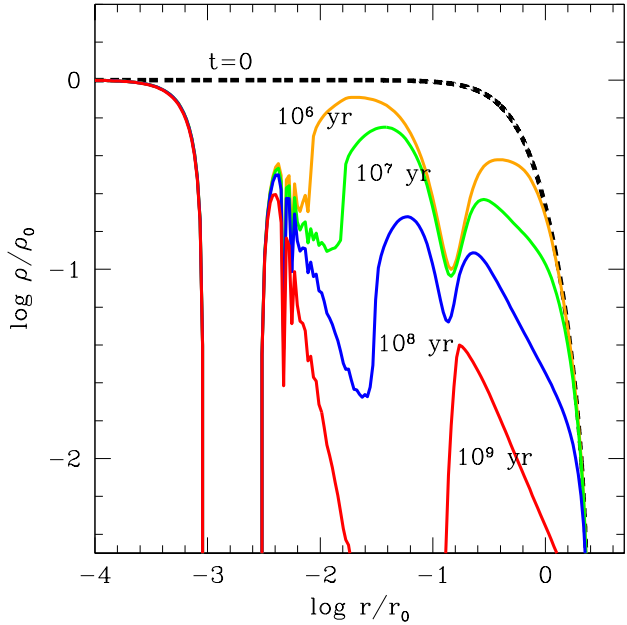}
\caption{
Radial stellar density profiles. 
Different colours show the stellar density at different evolution times. 
The time scales are calculated for a  star cluster with the mass of $10^{5}~\Msun$ and radius of $0.25~\rm pc$.
}
\label{fig:density}
\end{center}
\end{figure}

Following Equation~\ref{eq:mbh}, the stellar density at a specific radius decreases with time due to RR and NR processes.
Figure~\ref{fig:density} shows the time evolution of the  stellar density profile. 
Stars at $r \sim 10^{-3}~\rm pc$ have short RR relaxation time scales and are quickly consumed by the central BH, which in turn generates  a density gap. Stars close to the gap will be accreted next onto the BH.
%\adg{Then, stars at near outside the gap are accreted onto the BH in turn. }
Initially the radius at which the total enclosed stellar mass is smaller than the central BH, is limited to $r \lesssim 10^{-3}~\rm pc$ at $t \sim 10^{7}~\rm yr$, 
therefore, all stars at large radii are only  affected by NR. 
As the BH grows, the radius at which RR dominates becomes larger and reaches the radius of the cluster at $t \sim 10^{8}~\rm yr$.
Note that, however, we do not follow the change of radial position of stars in this work. 
The actual stellar distribution can become more smooth in detailed simulations \citep[e.g.,][]{Freitag02},
in which case the growth rate of the BH can be different.

\begin{figure}
\begin{center}
\includegraphics[scale=0.43]{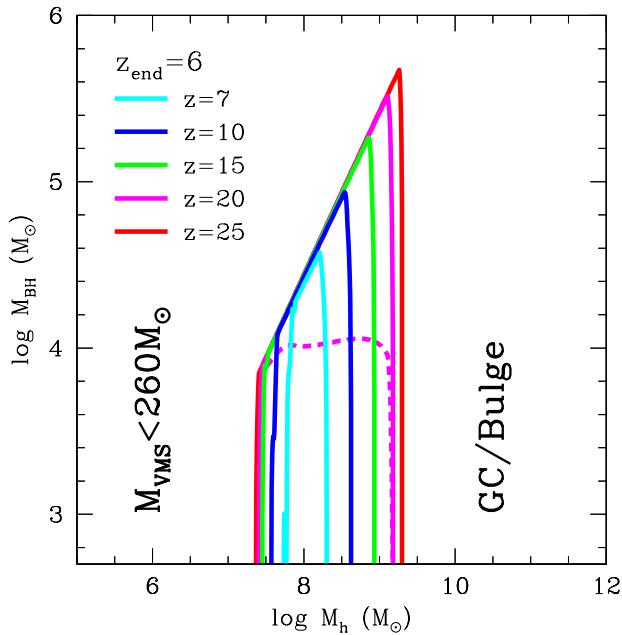}
\caption{
Black hole mass as a function of galaxy mass at $z = 6$. 
Different colour lines represent different redshifts at which galaxies merge. 
The magenta dashed-line is the black hole mass without 2-body relaxation for clusters formed at $z=20$.
}
\label{fig:mbh}
\end{center}
\end{figure}

We follow the time evolution of the BH growth from the merging event till redshifts $z=6$.
The distribution of  BH masses is shown in Figure~\ref{fig:mbh}.
As shown in Figure~\ref{fig:mvs}, only haloes in the limited mass range ($\Mh \sim 10^{7}-10^{9}~\Msun$)
contribute to the formation of MBH seeds. 

As equation~\ref{eq:NR} shows, the NR time scale is proportional to the stellar velocity dispersion and the inverse of stellar mass density. 
Since the stellar density is almost constant despite of the cluster's mass in our model, 
the BH growth rate is high even in low mass  haloes or stellar clusters.
However, as it will be show below, most of the stars even in massive star clusters accrete onto a central BH by $z=6$.
 Hence, the BH mass  linearly increases with halo mass.
We also compare the case without NR for galaxies that merge at  $z=20$, shown as the magenta dash line in the figure.
Unlike in the case of  both RR and NR processes active, 
the BH mass without NR is almost independent of halo mass,
and smaller than that with NR.
 
The radial dependence of the RR time scale implies that at small radii fast accretion of material onto the black hole will occur stalling the growth of the black hole until further material from larger radii is transported inwards. In the absence of NR the transport of material from larger radii is very inefficient in effect limiting the black hole mass in our model to $\sim 10^{4}~\Msun$. Thus NR is able to tap into the mass budget of stellar clusters at large radii and RR efficiently funnels this material close to the black hole for accretion.

\begin{figure}
\begin{center}
\includegraphics[scale=0.43]{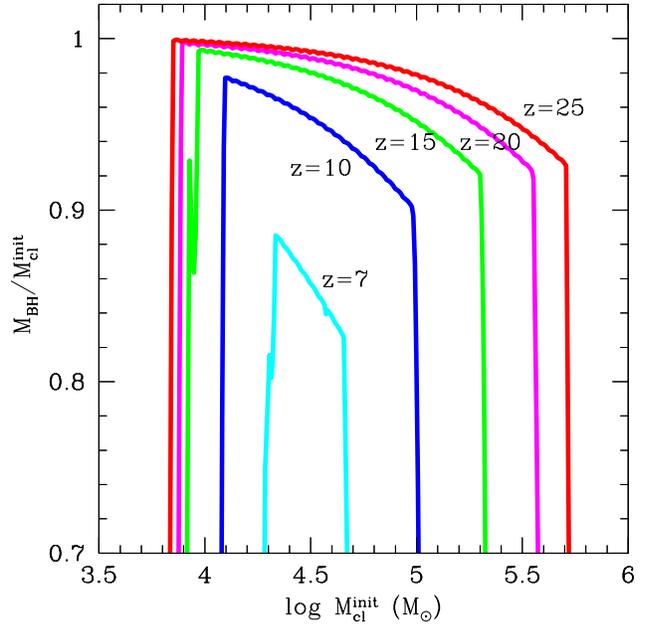}
\caption{
Final BH mass at $z=6$ normalized by the initial star cluster's mass.
Different colour lines represent different formation redshifts of the star clusters. 
}
\label{fig:mbhmst}
\end{center}
\end{figure}

Figure~\ref{fig:mbhmst} shows the conversion fraction of the initial stellar mass of a cluster into a MBH by $z=6$.
As shown in Figure~\ref{fig:rrtime}, the relaxation time scale is shorter than the average duration from the first merging event to $z=6$,
hence most of the stars ($\gtrsim$ ninety per cent) accrete onto a central BH.
With decreasing stellar density, the NR time scale gets longer, resulting in slow BH mass growth. 
The position at which the NR time scale becomes shortest gets closer to $R_{\rm cl}$ as the cluster mass decreases, 
resulting in smaller conversion factors of stars in  clusters with $\gtrsim 10^{5}~\Msun$.
In addition, as the stellar velocity dispersion increases with the star cluster's mass,
the NR time scale becomes longer, resulting in a somewhat inefficient conversion factor. 

\begin{figure}
\begin{center}
\includegraphics[scale=0.43]{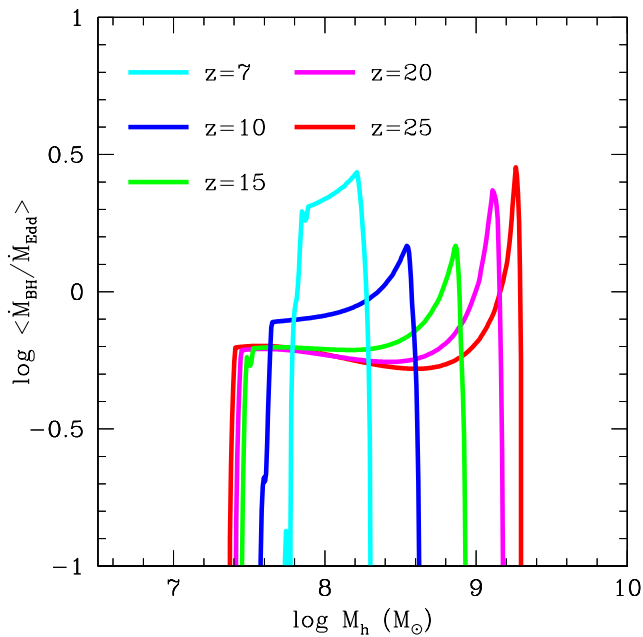}
\caption{
%Black hole mass as a function of galaxy mass at $z = 6$. 
Time-averaged stellar accretion rate normalized by the Eddington rate. 
Different color lines represent different redshifts at when galaxies merge. 
}
\label{fig:fedd}
\end{center}
\end{figure}

In order to study the efficiency of stellar mass accretion, 
we compare it with the Eddington rate. 
Figure~\ref{fig:fedd} represents the time averaged mass accretion rate normalized to the Eddington rate. 
The mass accretion rate increases with redshift, 
because  star clusters are more compact and the relaxation time scale becomes shorter for mergers at higher redshift in our model. 

Due to the short relaxation timescale, the clusters lose most of their stars before $z = 6$. 
After most of the stars are swallowed, the stellar accretion rate rapidly decreases as shown in Figure~\ref{fig:rrtime}.
 Therefore, the Eddington ratio at $z \ge 15$ is  lower than that at $z \le 10$.

However, even for star clusters formed at $z=25$, the time average accretion rate is close to the Eddington rate ($\dot{M}/\dot{M}_{\rm Edd} \sim 0.5$).
In addition, at $\Mh \sim 10^{9}~\Msun$, the accretion increases steeply. 
As shown in Figure~\ref{fig:mvs}, the initial BH (VMS) mass sharply drops at $\Mh \sim 10^{9}~\Msun$ due to the longer core-collapse timescale. 
On the other hand, the NR timescale is independent of the central BH mass. 
Hence, the Eddington ratio in the initial phase is very high.

Thus, we suggest that the stellar accretion can be the dominant mode for the growth of BHs up to a mass of $\sim 10^{5}~\Msun$.
For further growth of BHs to $\sim 10^{9}~\Msun$, efficient gas accretion is required.

%%%%Figure

%\begin{figure}
%\begin{center}
%\includegraphics[scale=0.43]{mbhgrow.eps}
%\caption{
%Black hole mass with evolution time in a star cluster
%with the mass of $10^{5}~\Msun$ and the radius of $0.4~\rm pc$.
%}
%\label{fig:sfr}
%\end{center}
%\end{figure}

\subsection{Number density of seeds of massive BHs}
To estimate the cosmological relevance of the above discussed formation channel, we combine our analytic model with a Monte-Carlo technique based dark matter merger tree.
%\adg{Based on the above results, we investigate statistical natures of MBHs. By combining merger history and halo mass function, we estimate number density of galaxies in major merging. Then, we derive black hole mass functions (BHMFs). }
In combination with the dark matter mass function at any given redshift this will allow to predict the merger rate of galaxies and the associated black hole mass function (BHMF).

Merger trees are constructed and applied based on the extended Press-Schechter formalism \citep{Lacey93} using the algorithm presented in \citet{Somerville99}. The merger trees have been successfully implemented and tested in previous work \citep[e.g.,][]{Khochfar09, Khochfar11}.
%\adg{We use an extended Press-Shechter formalism (Lacey 1993; Kochfar et al. 2006) to construct merger trees.} 
Our sample consists of 1000 merger tree realisations for each individual dark halo mass bin( $10^{8}~\Msun$ to $10^{13}~\Msun$.) Each tree is resolved down to a minimum halo mass of $10^{7}~\Msun$.  
We use cosmological parameters based on WMAP9 (Hinshaw et al. 2012), $\Omega_{\Lambda}=0.7$, $\Omega_{\rm M}=0.3$, $\sigma_{8}=0.9$ and $h=0.7$.

\begin{figure}
\begin{center}
\includegraphics[scale=0.43]{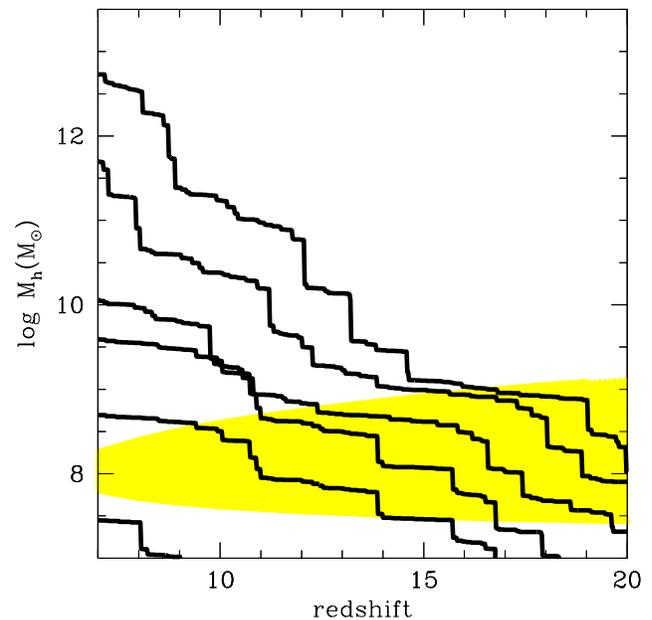}
\caption{
Example redshift evolution of haloes.
The yellow shaded region indicates the mass range in which haloes can form BH seeds when they experience major merging.
}
\label{fig:mergertree}
\end{center}
\end{figure}

Figure~\ref{fig:mergertree} shows the redshift evolution of six sampled haloes which reach the halo mass of $10^{8}, 10^{9}, 10^{10}, 10^{11}, 10^{12}$
and $10^{13}~\Msun$ at $z=6$ respectively. 
The yellow shade region represents the mass range in which the merging haloes can form MBH seeds. 
As seen in the Figure and expected, massive haloes experience mergers in the shaded region at higher redshifts. 
For example, most haloes with $\Mh \ge 10^{12}~\Msun$ at $z=6$ can form MBH seeds at $z \gtrsim 15$,
resulting in plausible seeds of SMBHs at $z \gtrsim 6$. 
In addition, due to the variation in the redhsift when the actual merger occurs, 
haloes with the same mass at $z=6$ can host MBHs of different mass.
Note that we assume the metallicity of the merging galaxies is low and remains so during the merger, and that the mass loss of VMSs is not significant. 
If galaxies are metal-enriched to a high degree, the initial BH mass is likely to change. 

Major mergers with mass ratio $\; \lesssim 1: 3$ are effective means in transporting angular momentum and allowing the gas to collapse to the central region of a galaxy (e.g. Bois et al. 2010). The efficiency of this process decreases with increasing mass ratio and we will in the following focus only at major mergers   with mass ratio $< 1: 2$. We have checked that inclusion of minor mergers does not change the BHMF significantly.
We identify the first major merger event in the history of a halo as the BH seeding event.

\begin{figure}
\begin{center}
\includegraphics[scale=0.43]{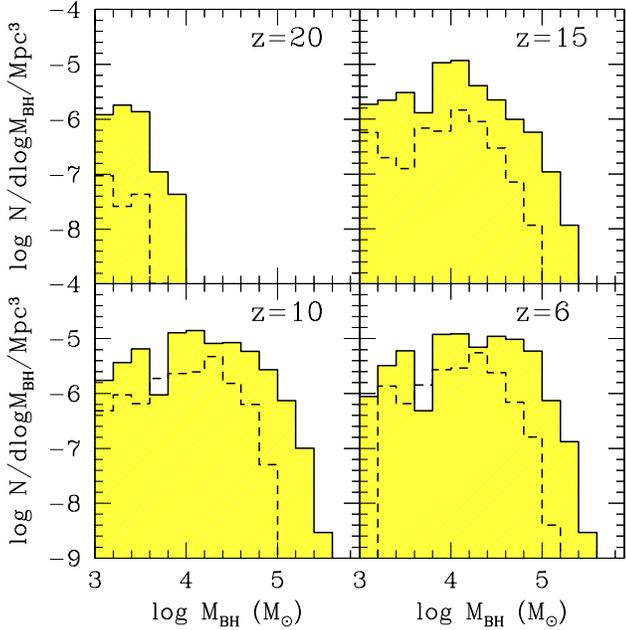}
\caption{
Black hole mass function at $z=20, 15, 10$ and $6$.
Solid and dash lines represent the case of $\epsilon_{\rm SF}=1$ (our fiducial model) and $0.3$,
where $\epsilon_{\rm SF}$ is the conversion efficiency from gas to stars in the inner region $r < r_{\rm cl}$.
}
\label{fig:bhmf}
\end{center}
\end{figure}

Figure~\ref{fig:bhmf} shows the predicted BHMFs at $z=6, 10, 15$ and $20$.
At $z = 20$, the BHMF is limited to $\MBH \lesssim 10^{4}~\Msun$ 
because the merging events just took place. As time progresses the number density of BHs and their average mass increases.
%\adg{The number density increases with decreasing redshift, and BHMFs have tails to higher BH mass.} 
%At $z \sim 10-15$ we find MBHs of $\sim 10^{5}~\Msun$ with  number densities of $\sim 10^{8-9}~\rm Mpc^{-3}$. 
%These number densities are similar to that of observed QSOs at $z \gtrsim 6$ \citep[e.g.,][]{Fan06b}.

At $z \sim 10-15$ we find MBHs of $\sim 10^{5}~\Msun$ with  number densities of $\sim 10^{-6}$ to $ 10^{-7}~\rm Mpc^{-3}$.
If we assume $\Mcl = 0.3 M_{\rm inf}$, the number densities decrease to $\sim 10^{-8}$ to $10^{-9}~\rm Mpc^{-3}$ 
which are similar to that of observed QSOs at $z \gtrsim 6$ \citep[e.g.,][]{Fan06b}. 
This is because the core-collapse time scale gets longer, hence the halo mass range becomes narrower. 
In addition, the growth rate of MBHs by stellar relaxation also becomes lower.

 MBHs of $10^{5}~\rm \Msun$ that grow close to the Eddington limit will reach $\sim 10^{9}~\Msun$ by $z \sim 6 - 7$ and thus could be the potential progenitors.
 To illustrate this point, one can consider a  MBH formed at $z=15 \;(10)$, it can reach  $10^{9}~\Msun$ at $z=6$ with an average Eddington rate of $0.6\; (0.9)$.
 
Cosmological simulations show that $10^{5}~\Msun$ MBHs at galactic centres can grow to $\sim 10^{9}~\Msun$ by $z \sim 6$ 
via gas accretion using a the Bondi-Hoyle model at the sub-grid level \citep[e.g,][]{DiMatteo05, DiMatteo12, Li07}. 
These simulations show that once a MBH is formed at the centre of galaxies in the halos with $M_{\rm h} \gtrsim 10^{10}~\Msun$, 
the gas keeps accreting onto the MBH close to the Eddington limit even under the presence of feedback \citep[see however,][for high resolution simulations including radiative transfer.]{Johnson11}.  
If we allow the MBH seeds of $10^{5}~\Msun$ to grow at the Eddington limit, the time needed to  grow from $10^{5}$ to $10^{9}~\Msun$ is
$t =\frac{\epsilon}{1-\epsilon} \tau_{\rm Sal}  {\rm  ln}10^{4} \sim  0.46~\rm Gyr$ for $\epsilon = 0.1$, 
where $\tau_{\rm Sal}$ is the Salpeter time scale $= \frac{c \sigma_{\rm T}}{4 \pi {\rm G} m_{\rm p}} $.
Hence, in order to explain SMBHs of $10^{9}~\Msun$ at $z \sim 6\; (7)$, MBH seeds of $10^{5}~\Msun$ are required at $z \gtrsim 10\; (13)$. 
Our models satisfy these constraint.

It is well known observationally that the mass of MBHs at galactic centres in local galaxies
tightly correlates with the stellar mass of galactic bulges \citep[e.g.,][]{Marconi03, Gultekin09}. 
On the other hand, at high redshifts, the BH-stellar mass relation is less well constraint, 
due to difficulties of observing BHs of $\lesssim 10^{8}~\Msun$.
If we compare our modeled BH mass density with recent observational results of stellar mass densities \citep{Stark13},
the BH mass density is approximately 5 orders of magnitude smaller than the observed stellar mass density at $z \sim 10$, and continues to become even smaller with respect to the stellar mass density. This implies that gas accretion potentially plays a dominant role in the growth of BHs after MBH seeds of $\sim 10^{5}~\Msun$ form.
%Here, using recent observational results of stellar mass densities (Stark et al. 2013), we compare our modelled BH mass density to the stellar mass density. Figure~\ref{fig:totalbh} shows the cumulative BH mass per unit volume. Due to the increase in the number density of haloes having  masses of $\sim 10^{7.5}-10^{9}~\Msun$ and the growth of BH masses, the cumulative BH mass increases with decreasing redshift. The predicted BH mass density is approximately 5 orders of magnitude smaller than the observed stellar mass density at $z \sim 10$, and continues to become even smaller with respect to the stellar mass density. This can be attributed to not including the effect of gas accretion onto our BH seeds in the model. High gas fractions in high-redshift galaxies mainly maintained by short cooling times are able to support star formation and black hole growth \citep[e.g.,][]{Khochfar11}.  As for the future evolution, any scatter in the BH mass bulge mass relation at early times will reduce and tend toward the observed relation at $z=0$ due to continued mergers of galaxies and the central-limit theorem \citep{Peng07, Hirschmann10}.

%\begin{figure}
%\begin{center}
%\includegraphics[scale=0.43]{sngrav.eps}
%\caption{
%Ratio of supernovae energy of initial starburst in star clusters ($\Esn$) to gravitational binding energy of gas disc ($\Eg$). 
%The dash line shows $\Esn = \Eg$. If $\Esn/\Eg > 1$, gas in disc can be evacuated or gas inflow to galactic centers can be suppressed due to
%supernovae feedback. 
%}
%\label{fig:sfr}
%\end{center}
%\end{figure}

%\begin{figure}
%\begin{center}
%\includegraphics[scale=0.43]{totalbh.eps}
%\caption{
%Mass density of stars or black holes in a comoving unit volume of $\rm Mpc^{3}$ as a function of redshift. 
%Square and circle symbols show observed stellar mass densities by \citet{Stark13} and \citet{Oesch14} respectively.
%Dashed lines show black hole masses derived by multiplying $10^{-3}$ to the observed stellar mass densities.
%The solid line is our result.
%}
%\label{fig:totalbh}
%\end{center}
%\end{figure}

%\begin{figure*}
%\begin{center}
%\includegraphics[scale=0.5, angle=90]{model.ps}
%\caption{
%in prep.
%}
%\label{fig:sfr}
%\end{center}
%\end{figure*}

%----------------------------------------------------------------------
%
% Section 4:  Discussion
%
%----------------------------------------------------------------------

\section{Discussion} 
\label{sec:discussion}

\subsection{Population III stars}
\begin{figure}
\begin{center}
\includegraphics[scale=0.43]{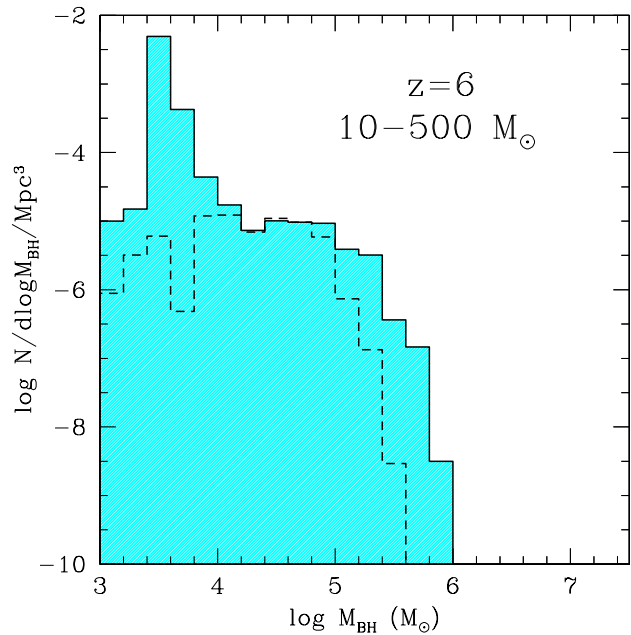}
\caption{
Black hole mass function at $z=6$
assuming  a top heavy IMF in the mass ranges $10$ to $500~\Msun$
with slope $\alpha=-2.35$.
Dash lines show our fiducial model ($M=0.1 - 100~\Msun$) as shown in Figure~\ref{fig:bhmf}.
}
\label{fig:popiii}
\end{center}
\end{figure}

In this work, we consider a Salpeter IMF in the mass range $0.1 - 100~\Msun$. 
However, the metallicity of gas is low at high redshift, and hence the IMF can tend to be top-heavy. 
As an upper limit, we consider zero metalicity stars, i.e., Population III (Pop III) stars.
The IMF of Pop III stars is still unknown due to missing observational probes as well as uncertainties in the modelling of their formation. Given that accretion rates onto proto-stellar seeds in primoridal gas clouds are high it is generally argued that Pop III stars should be more massive than their metal enriched counter parts 
\citep[e.g.,][]{Omukai98, Omukai03,  Nakamura01, Bromm02, Abel02a, Yoshida06, Yoshida08, Turk09, Stacy10, Stacy12, Clark11, Umemura12, Hirano14, Susa13, Susa14}.
% \adg{On the other hand, if first galaxies still have pristine gas, Population (Pop) III stars can form. Pop III stars tend to be massive because of higher gas accretion rate onto primordial stars (Bromm et al. 2001; Abel et al. 2001; Nakamura\&Umemura 2001; Yoshida et al. 2006; Yoshida et al. 2008; Turk et al. 2009; Clark et al. 2012; Umemura et al. 2012; Hirano et al. 2013; Susa et al. 2014).}
We re-examine our model trying to account for such massive Pop III stars assuming a top heavy IMF from  $10 - 500~\Msun$ \citep{Hirano14} and a slope $\alpha = - 2.35$. 
Pop III stars usually form in mini-haloes of $\sim 10^{6}~\Msun$ \citep[e.g.,][]{Bromm02}, 
as a single star \citep{Yoshida08} or binary \citep{Turk09} or multiple \citep{Clark11}. 
On the other hand, UV radiation in the  Lyman-Werner band from star forming galaxies can suppress star formation in neighbouring haloes \citep{Johnson13}. 
As a result, there may be the possibility that atomic cooling haloes host massive pristine gas \citep[e.g.,][]{Agarwal14}. 
%Recent observation detected He II line from a $\rm Ly\alpha$ emitting galaxy which suggested Pop III star cluster (Sobral et al. 2015).
The resulting BHMF at $z=6$ is shown in Figure~\ref{fig:popiii}. 
In contrast to star clusters with a fiducial IMF star clusters with a top-heavy IMF produce more massive black holes at their centre per unit mass in stars. 
%\adg{In the case of the top heavy IMF, all haloes leave BH seeds because of the maximum mass stars directly collapse into the BHs even without the growth by the core-collapse, while the mass range for the normal IMF is limited as shown in Figure~\ref{fig:mvs}.}
Massive Pop III stars cause core-collapse of more massive haloes within $3~\rm Myr$, and
lead to the formation of heavier VMSs. 
In addition,
because of the increase of the average stellar mass $<m>$ from $0.3~\Msun$ for the normal IMF to $30~\Msun$ for the Pop III IMF, 
the relaxation time scale becomes shorter by factor $\sim 10$ (Equation~\ref{eq:NR} and \ref{eq:RR}). 
As a result, massive BHs of $\sim 10^{6}~\Msun$  can form by $z \sim 6$. 

In other words, for such a top heavy IMF, even star clusters with similar size and density of typical globular clusters in the local universe
can cause core-collapse, and form VMSs.

%\adg{In addition, due to the higher maximum mass of stars, more massive star-clusters do the core-collapse within 3 Myr. Therefore, even massive large star clusters in massive galaxies go to the system consisting of central BHs and stars,and then, the BHs grow up. }

Note that, however, that massive galaxies ($\gtrsim 10^{8}~\Msun$) are likely metal enriched due to type-II SNe of prior stars \citep[e.g.,][]{Maio11, Wise12b}.
Therefore, massive star clusters consisting of Pop III stars may not form in practice. 
On the other hand, if metal mixing is not efficient, some pockets of pristine gas will be still able to lead to the formation of Pop III star clusters, with short core-collapse time scales \citep[see however,][]{Smith15}. 

Local metal poor globular clusters can give constraints on the stellar IMF, density and the cluster's size
to explain their old-age of $\gtrsim 10$ Gyr.
Based on our results star clusters with a local IMF and  initial size of $\gtrsim 1~\rm pc$ should be long-lived.

\subsection{Metal poor globular clusters}

%\adg{As discussed in Section~\ref{sec:result}, merging galaxies of $\Mh \sim 10^{9} -- 10^{10}~\Msun$ 
%can make globular clusters at the galactic centers. 
%Here, we estimate \adb{the initial mass function} of globular clusters by considering galaxies with $\tcc > 3~\rm Myr$ and $\Esn > \Eg$. }
Merging haloes between $\Mh \sim 10^{9} - 10^{10}~\Msun$ in our model result in the formation of globular clusters. 
We here present the mass function focusing on those with $\tcc > 3~\rm Myr$ and $\Esn > \Eg$. 
For the star clusters with $\tcc > 3~\rm Myr$, central massive BH seeds via  core-collapse do not form, leaving the stars unaffected in the cluster. 
In addition, if $\Esn > \Eg$, SN feedback is likely to suppress further gas accretion onto the star clusters, 
resulting in isolated systems that are dense and compact.
It is widely known that observed globular clusters do not have dark matter haloes. 
On the other hand, some theoretical models showed the formation of globular clusters in dark matter haloes, 
and the number density nicely fit with observations \citep{Moore06, Trenti15, Kimm15}.
These authors assumed the hosting dark matter haloes would subsequently be stripped away by tidal interaction with galaxies. 
Although the formation mechanism of globular cluster is still unclear, 
we here present the predicted mass distribution from our merger model as a possible sub-dominant channel to forming globular clusters. 

\begin{figure}
\begin{center}
\includegraphics[scale=0.43]{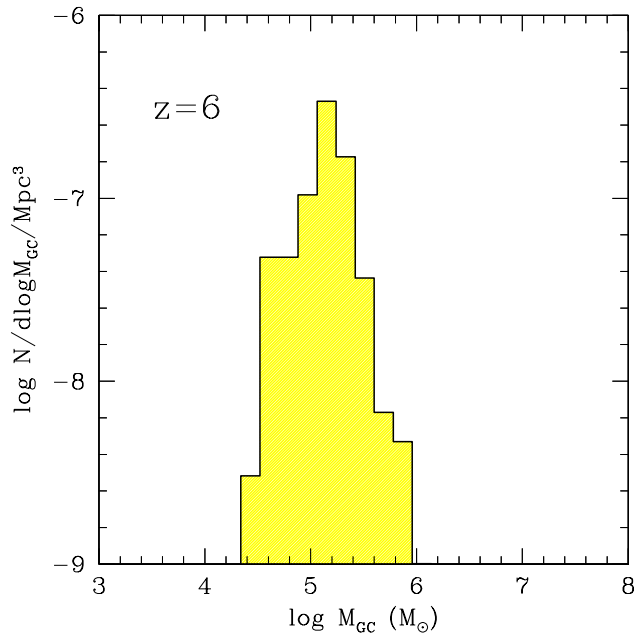}
\caption{
Mass function of globular clusters resulting from mergers of galaxies. 
}
\label{fig:gcmf}
\end{center}
\end{figure}

Figure~\ref{fig:gcmf} shows the mass function at $z = 6$.
The mass function has a peak at $M \sim 1.5 \times 10^{5}~\Msun$ and the mass range is consistent with that of observed globular clusters in local Universe. Recent observation of metal-poor globular clusters (MPGCs) show that the stellar population is very old, and compatible with formation at $z \gtrsim 6$ \citep{Brodie06} % \& Strader 2006).
%\adg{In addition, some fraction of MPGCs showed very old which is similar to cosmic age. Therefore, some of them are likely to form in first galaxies at $z \gtrsim 6$.} 
We compare the number density of MPGCs in our model with that of the observed one in local Universe. 
%\sout{
%Figure~\ref{fig:gcrate} shows the redshift evolution of MPGCs. 
%The number density increases with decreasing redshift, and reaches $\Ngc \sim 5 \times 10^{-7}~\rm Mpc^{-3}$ at $z = 6$ similar to the local number density  of  MPGCs which is  $\Ngc \sim 1 ~\rm Mpc^{-3}$ }
%\citep{Barmby00, Forbes00}.
Thus, our model alone cannot reproduce the number density of the observed local MPGCs, while it matches their physical properties (size and mass).

%\adg{On the other hand, recent observations of local MPGCs suggested the $\Ngc \sim 1 ~\rm Mpc^{-3}$. Therefore, merging first galaxies can be a mechanism of MPGC formation, however, the number fraction is much smaller than the observed one because of the rarity of major merging processes. } 
%\adg{A large fraction of MPGCs show old age ($\gtrsim 10~\rm Gyr$), hence some of them should form in first galaxies. More general process of MPGC formation in local region would be required to explain the observed number density. } \adr{SK: you have said this already earlier}

The subsequent interaction with their environment via processes like tidal stripping and merging \citep[e.g,][]{Leon00}  will change the mass function of  MPGCs and its final outcome will depend on the future evolution.
%\adg{In addition, our results suggest the ``initial'' mass of MPGCs. The mass of MPGCs can decreases due to tidal stripping. Hence, the PDF can shift to lower mass with time. In addition, with growth of galaxies via merging, some fraction of MPGCs can be incorporated into bulges. The evolution of the distribution and mass of MPGCs hence depends on the growth history of galaxies and the environments. }

Recently \citet{Trenti15} modeled local MPGCs in cosmological N-body simulations \citep[see also][]{Moore06}.
They assumed GCs form in merging primordial haloes of $10^{8}~\Msun$ at $z \sim 10$, and used their model to explain the observed nature of local MPGCs, 
e.g., age and metallicity distribution, and their spacial distribution.
Our model of GC formation requires somewhat more massive merging haloes ($\gtrsim 10^{9}~\Msun$) suggesting a possible halo mass bias leading to the formation of black hole seeds.

%\begin{figure}
%\begin{center}
%\includegraphics[scale=0.43]{gcrate.eps}
%\caption{
%Redshift evolution of the number density of globular clusters. 
%}
%\label{fig:gcrate}
%\end{center}
%\end{figure}

%----------------------------------------------------------------------
%
% Section 5:  Summary
%
%----------------------------------------------------------------------

\section{Summary}
\label{sec:summary}
In this work, we analytically model the formation of MBHs in the first merging galaxies. These BHs provide the natural seeds for SMBH at $z \sim 6$. 
 We show that compact star clusters that likely form during such mergers are prone to core-collapse and produce VMSs of $\sim 1000~\Msun$ at their centres.
%\adg{formed in merging galaxies due to rapid gas inflow to galactic centres. The star clusters cause core-collapse and make VMSs of $\sim 1000~\Msun$ at the centers.}
VMSs collapse to BHs without going through a SN stage and subsequently grow via swallowing stars in their vicinity. Relaxation processes  are efficient in refilling the loss cone around the BH continuously with new stars.  
%\adg{The VMSs result in BHs without SNe. Thereafter the BHs grow up via through swallowing residual stars in the clusters with angular momentum transport by stellar relaxation processes. In particular, resonant relaxation, process help to re-fill loss cone in stellar orbits, and enhances the growth of BHs.} 
We find that the BHs mass  sensitively depends on the 
the radius and mass of the hosting star clusters which determine  the core-collapse and stellar relaxation time scales.
Within this scenario major mergers of galaxies with $\gtrsim 4\times10^{8}~\Msun$ at $z \sim 20$ lead to the formation of $\gtrsim 10^{5}~\Msun$ BHs by $z \sim 10$ which are likely progenitors of  SMBHs at $z \gtrsim 6$.
%As a result, major merging of galaxies of $\gtrsim 4\times10^{8}~\Msun$ at $z \sim 20$ successfully leads to the formation of $\gtrsim 10^{5}~\Msun$ BHs by $z \sim 10$which can be seeds of SMBHs at $z \gtrsim 6$.

Based on our results the average relation between host stellar mass and black hole mass will not be the same as observed in the local universe. 
After the formation of first massive BHs, the BHs can grow by gas accretion, and their mass can be close to the locally observed relation \citep{DiMatteo05}.
In addition, such deviation will also
reduce over time as merging of galaxies and black holes will move host galaxies and black holes closer to the locally observed relation as a consequence of the central limit theorem \citep{Peng07, Hirschmann10}.

Our model predictions are sensitive to the properties of the hosting star clusters, e.g., mass, radius, IMF of stars. We here assumed physically motivated analytic relations  that we will further test and systematically investigate in a follow-up study using detailed numerical simulations.   
%\adg{The evolution from massive gas clouds to star clusters has not been understood well yet. We will investigate the nature of star clusters in low-metal gas clouds in future work by detailed numerical simulations. }

%----------------------------------------------------------------------
%
% Acknowledge
%
%----------------------------------------------------------------------
\section*{Acknowledgments}
We are grateful to N. Yoshida and A. Ferrara for valuable discussion. We would also like to thank the referee for his/her careful reading of the manuscript.
%This work is supported in part by the NSF grant AST-0807491, 

%----------------------------------------------------------------------
%
% References
%
%----------------------------------------------------------------------
%\bibliographystyle{mn}

%\bibliography{mn-jour,HY}

\label{lastpage}

\end{document}